\documentstyle[twoside,fleqn,espcrc2,epsfig]{article}

\title{Quantum Spins and Quantum Links: \\
The D-Theory Approach to Field Theory}

\author{U.-J. Wiese 
\address{Center for Theoretical Physics, Laboratory for Nuclear Science and 
Department of Physics \\ 
Massachusetts Institute of Technology (MIT), Cambridge, MA 02139}}

\begin{document}

\begin{abstract}

A new non-perturbative approach to quantum field theory is proposed. Instead of
performing a path integral over configurations of classical fields, D-theory 
works with discrete quantized variables. Classical spin fields are replaced by 
quantum spins, and classical gauge fields are replaced by quantum links. The 
classical fields of a $d$-dimensional quantum field theory reappear as 
low-energy effective degrees of freedom of the discrete variables, provided the
$(d+1)$-dimensional D-theory is mass\-less. When the extent of the extra 
Euclidean dimension becomes small in units of the correlation length, an 
ordinary $d$-dimensional quantum field theory emerges by dimensional reduction.
The D-theory formulation of some spin models and gauge theories is constructed 
explicitly. In particular, QCD emerges as a quantum link model. 

\end{abstract}

\maketitle

\newpage

\section{Introduction}

Field theories are usually quantized by performing a path integral over
configurations of classical fields. This is the case both in perturbation 
theory and in Wilson's non-perturbative lattice formulation of quantum field
theory. However, there is another form of quantization, which is well-known
from quantum mechanics: a classical angular momentum vector can be replaced by
a vector of Pauli matrices. The resulting quantum spin is described by discrete
variables $\pm 1/2$, while the original classical angular momentum vector
represents a continuous degree of freedom. Here we propose to apply this kind
of quantization to field theory.

Of course, it is far from obvious that such a quantization procedure is 
equivalent to the usual one. For example, a single spin $1/2$ has the same
symmetry properties as a classical angular momentum vector, but it operates in 
a finite Hilbert space. How can the full Hilbert space of a quantum field 
theory be recovered when the classical fields are replaced by analogs of
quantum spins? Indeed, as we will see, this requires a specific dynamics, 
which, however, is generic in a wide variety of cases. This includes spin 
models and gauge theories, and, in particular, QCD. In these cases, a 
collective excitation of a large number of discrete variables acts as a 
classical field, in the same way as many spins $1/2$ can act as a classical 
angular momentum vector. To collect a large number of discrete variables, it 
turns out to be necessary to formulate the theory with an additional Euclidean 
dimension. When the $(d+1)$-dimensional theory is mass\-less, the classical 
fields emerge as low-energy excitations of the discrete variables. When the 
extent of the extra dimension becomes small in units of the correlation length,
the desired $d$-dimensional quantum field theory emerges via dimensional 
reduction. {\em Dimensional} reduction of {\em discrete} variables is a generic
phenomenon that occurs in a variety of models, thus leading to a new 
non-perturbative formulation of quantum field theory, which we call 
{\em D-theory}.

The chiral symmetries of fermions have already motivated extensions to five 
dimensions. In order to solve the lattice fermion doubling problem, Kaplan has 
proposed to view our 4-d space-time as a domain wall in five dimensions. A 5-d 
fermion, which is always vector-like, then develops a zero mode bound to the 
domain wall, which appears as a chiral fermion from a 4-d point of view. For 
QCD, Shamir has simplified Kaplan's ideas by working with fixed boundary
conditions for the fermions in a 5-d slab. In D-theory, the extra dimension 
appears for reasons related to the bosonic degrees of freedom. Still, as we 
will see, the bosonic D-theory construction fits very naturally with Shamir's
variant of Kaplan's fermion proposal.

D-theory has several interesting features that go beyond Wilson's 
non-perturba\-tive lattice formulation of quantum field theory. For example, 
due to the use of discrete variables, the theory can be completely fermionized.
All bosonic fields can be written as pairs of fermionic constituents, which we 
call rishons. The two indices of a bosonic matrix field --- for example, the 
two color indices of a gluon field matrix --- can be separated because they are
carried by two different rishons. This leads to new ways to attack the large 
$N$ limit of QCD and other interesting field theories. D-theory is attractive
also from a computational point of view. Discrete variables are particularly 
well suited for numerical simulations using very powerful cluster algorithms. 
In fact, the first cluster algorithm for a model with a continuous gauge group 
has been constructed for a $U(1)$ quantum link model, and it is plausible that 
this construction can be extended to non-Abelian gauge theories.

$SU(2)$ and $U(1)$ quantum link models were written down by Horn as early as
1981 \cite{Hor81}, and they were rediscovered and discussed in more detail by
Orland and Rohrlich in 1990 \cite{Orl90}. Recently, they have been rediscovered
again, and it has been realized how they are related to ordinary gauge theories
via dimensional reduction \cite{Cha97}. Quantum link QCD was constructed in
\cite{Bro97} and D-theory was discussed in \cite{Bea98}. A detailed analysis of
the $U(1)$ quantum link model is the subject of three contributions to these
proceedings \cite{Bea98a}.

This talk is organized as follows. In section 2, D-theory is explained in the
context of the $O(3)$ model. Section 3 contains the D-theory representation of 
complex vector and matrix fields. This leads to the explicit construction of 
some models in section 4. The dimensional reduction of quantum links models to 
ordinary gauge theories is discussed in section 5, and the inclusion of
fermions is described in section 6. Finally, section 7 contains the
conclusions.

\section{The $O(3)$ Model from D-Theory}
 
Let us illustrate the basic ideas behind D-theory in the simplest example
--- the 2-d $O(3)$ model, which we view as a Euclidean field theory in $1+1$
dimensions. Like QCD, this model is asymptotically free, and has a 
non-perturbatively generated mass-gap. In Wilson's formulation of lattice field
theory the model is formulated in terms of classical 3-component unit vectors 
$\vec s_x$ located on the sites $x$ of a quadratic lattice. The Euclidean 
action of the model is given by
\begin{equation}
S[\vec s] = - \sum_{x,\mu} \vec s_x \cdot \vec s_{x+\hat\mu},
\end{equation}
where $\hat\mu$ represents the unit vector in the $\mu$-direction. The theory
is quantized by considering the classical partition function
\begin{equation}
Z = \int {\cal D}\vec s \ \exp(- \frac{1}{g} S[\vec s]),
\end{equation}
which represents a path integral over all classical spin field configurations 
$[\vec s]$. Here $g$ is the coupling constant. Due to asymptotic freedom, the
continuum limit of the lattice model corresponds to $g \rightarrow 0$. In this
limit the correlation length $\xi \propto \exp(2 \pi/g)$ diverges 
exponentially. The strength of the exponential increase is given by the 1-loop 
$\beta$-function coefficient $2 \pi$ of the 2-d $O(3)$ model.

In contrast to the standard procedure, in D-theory one does not quantize by 
integrating over the classical field configurations $[\vec s]$. Instead, each 
classical vector $\vec s_x$ is replaced by a quantum spin operator $\vec S_x$ 
(a Pauli matrix for spin 1/2) with the usual commutation relations
\begin{equation}
[S_x^i,S_y^j] = i \delta_{xy} \epsilon_{ijk} S_x^k.
\end{equation}
The classical action of the 2-d $O(3)$ model is replaced by the Hamilton 
operator
\begin{equation}
H = J \sum_{x,\mu} \vec S_x \cdot \vec S_{x+\hat\mu},
\end{equation}
thus, leaving us with a quantum Heisenberg model. Here we restrict ourselves to
antiferromagnets, i.e., to $J>0$. Ferromagnets have a conserved order 
parameter, and therefore require a special treatment. Like the classical action
$S[\vec s]$, the Hamilton operator $H$ is invariant under global $SO(3)$ 
transformations. In quantum mechanics this follows from $[H,\vec S] = 0$, where
\begin{equation}
\vec S = \sum_x \vec S_x
\end{equation}
is the total spin. D-theory is defined by the quantum partition function
\begin{equation}
Z = \mbox{Tr} \exp(- \beta H).
\end{equation}
The trace is taken in the Hilbert space, which is a direct product of the
Hilbert spaces of individual spins. It should be noted that D-theory can be
formulated with any value of the spin, not only with spin 1/2.

At this point, we have replaced the 2-d $O(3)$ model, formulated in terms of 
classical fields $\vec s_x$, by a 2-d system of quantum spins $\vec S_x$ with 
the same symmetries. The inverse temperature $\beta$ of the quantum system can 
be viewed as the extent of an additional third dimension. In the standard 
interpretation of the 2-d quantum spin system this dimension would be Euclidean
time. In D-theory, however, Euclidean time is already part of the 2-d lattice.
Indeed, as we will see, the additional Euclidean dimension ultimately 
disappears via dimensional reduction. The 2-d antiferromagnetic spin 1/2 
quantum Heisenberg model has very interesting properties. It describes the 
undoped precursor insulators of high-temperature superconductors --- materials 
like $\mbox{La}_2\mbox{CuO}_4$ --- whose ground states are N\'eel ordered with 
a spontaneously generated staggered magnetization. Indeed, there is 
overwhelming numerical evidence that the ground state of the 2-d 
antiferromagnetic spin 1/2 quantum Heisenberg model exhibits long-range order 
\cite{Bar91,Wie94,Bea96}. The same is true for higher spins, and thus the 
following discussion applies equally well to all spin values. In practice, 
however, the smallest spin 1/2 is most interesting, because it allows us to 
represent the physics of the 2-d $O(3)$ model in the smallest possible Hilbert 
space.

Formulating the 2-d quantum model as a path integral in the extra dimension 
results in a 3-d $SO(3)$-symmetric classical model. At zero temperature of the 
quantum system we are in the infinite volume limit of the corresponding 3-d 
model. The N\'eel order of the ground state of the 2-d quantum system implies 
that the corresponding 3-d classical system is in a broken phase, in which only
an $SO(2)$ symmetry remains intact. As a consequence of Goldstone's theorem, 
two mass\-less bosons arise --- in this case two antiferromagnetic magnons (or 
spin-waves). Using chiral perturbation theory one can describe the magnon 
dynamics at low energies \cite{Leu90}. The Goldstone bosons are represented by 
fields in the coset $SO(3)/SO(2) = S^2$, which resembles a 2-dimensional 
sphere. Consequently, the magnons are described by 3-component unit vectors 
$\vec s$ --- the same fields that appear in the original 2-d $O(3)$ model. 
Indeed, due to spontaneous symmetry breaking, the collective excitations of 
many discrete quantum spin variables form an effective continuous classical 
field $\vec s$. This is one of the main dynamical ingredients of D-theory. 

Another ingredient is dimensional reduction, to which we now turn. To lowest 
order in chiral perturbation theory, the effective action of the Goldstone 
bosons takes the form
\begin{equation}
\label{spinaction}
S[\vec s] = \int_0^\beta dx_3 \int d^2x \ \frac{\rho_s}{2}
[\partial_\mu \vec s \cdot \partial_\mu \vec s + 
\frac{1}{c^2} \partial_3 \vec s \cdot \partial_3 \vec s].
\end{equation}
Here $c$ and $\rho_s$ are the spin-wave velocity and the spin stiffness. Note
that $\mu$ extends over the physical space-time indices 1 and 2 only. The 2-d 
quantum system at finite temperature corresponds to a 3-d classical model with 
finite extent $\beta$ in the extra dimension. For mass\-less particles --- 
i.e., in the presence of an infinite correlation length $\xi$ --- the finite
temperature system appears dimensionally reduced to two dimensions, because 
$\beta \ll \xi$. In two dimensions, however, the Mermin-Wagner-Coleman theorem 
prevents the existence of interacting mass\-less Goldstone bosons \cite{Mer66},
and, indeed, the 2-d $O(3)$ model has a non-perturbatively generated mass-gap. 
Hasenfratz and Niedermayer suggested to use a block spin renormalization group 
transformation to map the 3-d $O(3)$ model with finite extent $\beta$ to a 2-d 
lattice $O(3)$ model \cite{Has91}. One averages the 3-d field over volumes of 
size $\beta$ in the third direction and $\beta c$ in the two space-time 
directions. Due to the large correlation length, the field is essentially 
constant over these blocks. The averaged field is defined at the block centers,
which form a 2-d lattice of spacing $\beta c$. Note that this lattice spacing
is different from the lattice spacing of the original quantum antiferromagnet.
The effective action of the averaged field defines a 2-d lattice $O(3)$ model,
formulated in Wilson's framework. Using chiral perturbation theory, Hasenfratz 
and Niedermayer expressed its coupling constant as
\begin{equation}
1/g = \beta \rho_s + {\cal O}(1/\beta \rho_s).
\end{equation}
Using the 3-loop $\beta$-function of the 2-d $O(3)$ model together with its
exact mass-gap \cite{Has90}, they also extended an earlier result of 
Chakravarty, Halperin and Nelson \cite{Cha89} for the inverse correlation 
length of the quantum antiferromagnet to
\begin{equation}
\xi \! = \! \frac{ec}{16 \pi \rho_s} \exp(2 \pi \beta \rho_s)
[1 - \frac{1}{4 \pi \beta \rho_s} + {\cal O}(1/\beta^2 \rho_s^2)].
\end{equation}
Here $e$ is the base of the natural logarithm. The above equation resembles
the asymptotic scaling behavior of the 2-d classical $O(3)$ model. In fact, one
can view the 2-d antiferromagnetic quantum Heisenberg model in the zero 
temperature limit as a regularization of the 2-d $O(3)$ model. It is remarkable
that this D-theory formulation is entirely discrete, even though the model is 
usually formulated with a continuous classical configuration space.

\begin{figure}[t]
\epsfig{figure=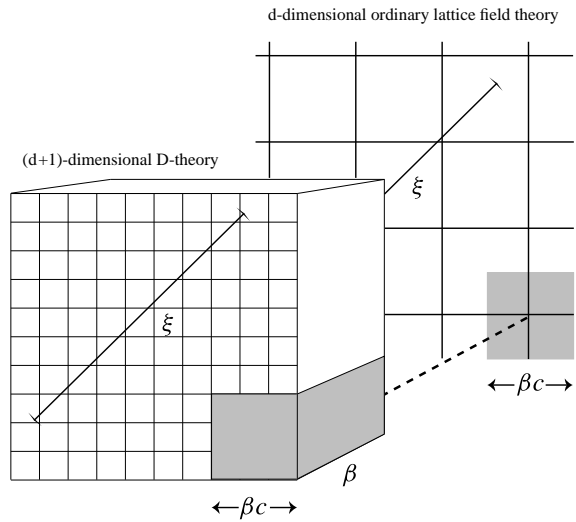,width=7.5cm}
\vspace*{-1cm}
\caption{Dimensional reduction of a D-theory: Averaging the $(d+1)$-dimensional
effective field of the D-theory over blocks of size $\beta$ in the extra 
dimension and $\beta c$ in the physical directions results in an effective
$d$-dimensional Wilsonian lattice field theory with lattice spacing $\beta c$.}
\vspace{-0.5cm}
\label{dtheory}
\end{figure}

As illustrated in figure \ref{dtheory}, the dimensionally reduced effective 2-d
theory is a Wilsonian lattice theory with lattice spacing $\beta c$.
The continuum limit of that theory is reached as $g = 1/\beta \rho_s 
\rightarrow 0$, and hence as the extent $\beta$ of the extra dimension becomes
large. Still, in physical units of the correlation length, the extent $\beta
\ll \xi$ becomes negligible in this limit, and hence the theory undergoes 
dimensional reduction to two dimensions. In the continuum limit, the lattice
spacing $\beta c$ of the effective 2-d Wilsonian lattice $O(3)$ model becomes 
large in units of the microscopic lattice spacing of the quantum spin system.
Therefore, D-theory introduces a discrete substructure underlying Wilson's
lattice theory. This substructure is defined on a very fine microscopic 
lattice. In other words, D-theory regularizes quantum fields at much shorter
distance scales than the ones considered in Wilson's formulation. For practical
purposes it is essential that the effective Wilsonian lattice gauge theory
results from exact blocking of the continuum field theory describing the
low-energy excitations of the underlying D-theory. Hence, the resulting 
Wilsonian lattice action is perfect, up to cut-off effects due to the 
microscopic D-theory lattice, such that in D-theory simulations lattices need
not be finer than in the standard approach.

The additional microscopic structure may provide new insight into the 
long-distance continuum physics. In the context of the quantum Heisenberg 
model, the microscopic substructure is due to the presence of electrons hopping
on a crystal lattice. After all, the spin-waves of a quantum antiferromagnet 
are just collective excitations of the spins of many electrons. In the same 
way, gluons appear as collective excitations of rishons hopping on the 
microscopic lattice of the corresponding quantum link model for QCD. In that 
case, the lattice is most likely unphysical, because it just serves as a 
regulator. However, even if the rishons propagate only at the cut-off scale, 
they may still be useful for understanding the physics in the continuum limit. 
Let us illustrate the rishon ideas in the context of the quantum Heisenberg 
model. Then the rishons can be identified with physical electrons. In fact, the
quantum spin operator at a lattice site $x$,
\begin{equation}
\vec S_x = \frac{1}{2} \sum_{i,j} c_x^{i\dagger} \vec \sigma_{ij} c_x^j,
\end{equation}
can be expressed in terms of Pauli matrices $\vec \sigma$ and electron creation
and annihilation operators $c_x^{i\dagger}$ and $c_x^i$ ($i,j \in \{1,2\}$)
with the usual anti-commutation relations
\begin{equation}
\{c_x^{i\dagger},c_y^{j\dagger}\} = \{c_x^j,c_y^j\} = 0, \ 
\{c_x^i,c_y^{j\dagger}\} = \delta_{x,y} \delta_{ij}.
\end{equation}
It is straightforward to show that $\vec S_x$, constructed in this way, has the
correct commutation relations. In fact, the commutation relations are also
satisfied when the rishons are quantized as bosons. The rishon representation
allows us to rewrite the Hamilton operator in terms of singlet combinations
$\sum_i c_x^{i\dagger} c_y^i$, with interesting consequences for $O(N)$ quantum
spin models in the large $N$ limit. It should be noted that the total number of
rishons at each site $x$ is a conserved quantity, because the local rishon 
number operator ${\cal N}_x = \sum_i c_x^{i\dagger} c_x^i$ commutes with the 
Hamiltonian. In fact, fixing the number of rishons is equivalent to selecting a
value for the spin, i.e., to choosing an irreducible representation.

The discrete nature of the D-theory degrees of freedom allows the application 
of very efficient cluster algorithms. The quantum Heisenberg model, for 
example, can be treated with a loop cluster algorithm \cite{Eve93,Wie94}. 
Defining the path integral for discrete quantum systems does not even require 
discretization of the additional Euclidean dimension. This observation has led 
to a very efficient loop cluster algorithm operating directly in the continuum 
of the extra dimension \cite{Bea96}. This algorithm, combined with a 
finite-size scaling technique, has been used to study the correlation length
of the Heisenberg model up to $\xi \approx 350000$ lattice spacings 
\cite{Bea97}. In this way the analytic prediction of Hasenfratz and Niedermayer
--- and hence the scenario of dimensional reduction --- has been verified 
numerically. This shows explicitly that the 2-d O(3) model can be investigated 
very efficiently using D-theory, i.e., by simulating the (2+1)-d path integral 
for the 2-d quantum Heisenberg model. In this case, the numerical effort is 
compatible to simulating the 2-d $O(3)$ model directly with the Wolff cluster 
algorithm \cite{Wol89}. However, for most other lattice models --- for example,
for gauge theories --- despite numerous attempts, no efficient cluster 
algorithm has been found in Wilson's formulation. If an efficient cluster 
algorithm can be constructed for the D-theory formulation, it would allow 
simulations more accurate than the ones presently possible. Indeed, recently, 
the first cluster algorithm for a model with a continuous gauge group has been 
constructed for a $U(1)$ quantum link model \cite{Bea98}, and it is plausible 
that it can be generalized to non-Abelian gauge groups.

The exponential divergence of the correlation length is due to the asymptotic
freedom of the 2-d $O(3)$ model. Hence, one might expect that the above 
scenario of dimensional reduction is specific to $d=2$. As we will see now,
dimensional reduction also occurs in higher dimensions, but in a slightly
different way. Let us consider the antiferromagnetic quantum Heisenberg model 
on a $d$-dimensional lattice with $d>2$. Then, again, the ground state has a
broken symmetry, and the low energy excitations of the system are two 
mass\-less magnons. The effective action of chiral perturbation theory is the 
same as before, except that the integration now extends over a 
higher-dimensional space. Again, at an infinite extent $\beta$ of the extra 
dimension, one has an infinite correlation length. Thus, once $\beta$ becomes 
finite, the extent of the extra dimension is negligible compared to the 
correlation length, and the theory undergoes dimensional reduction to $d$ 
dimensions. However, in contrast to the $d=2$ case, now there is no reason why 
the Goldstone bosons should pick up a mass after dimensional reduction. 
Consequently, the correlation length remains infinite, and we end up in a 
$d$-dimensional phase with broken symmetry. When the extent $\beta$ is reduced 
further, we eventually reach the symmetric phase, in which the correlation 
length is finite. The transition between the broken and symmetric phase is 
known to be of second order. Thus, approaching the phase transition from the 
symmetric phase at low $\beta$ also leads to a divergent correlation length, 
and hence, again, to dimensional reduction. Thus, the universal continuum 
physics of $O(3)$ models in any dimension $d \geq 2$ is naturally contained in 
the framework of D-theory.

Still, the $d=1$ case requires a separate discussion. The behavior of quantum 
spin chains depends crucially on the value of the spin. Haldane has conjectured
that 1-d antiferromagnetic $O(3)$ quantum spin chains with integer spins have a
mass-gap, while those with half-integer spins are gapless \cite{Hal83}. This
conjecture is by now verified in great detail. For example, the spin 1/2 
antiferromagnetic Heisenberg chain has been solved by the Bethe ansatz, and 
indeed turns out to have no mass-gap \cite{Bet31}. The same has been shown 
analytically for all half-integer spins \cite{Lie61}. On the other hand, there 
is strong numerical evidence for a mass-gap in spin 1 and spin 2 systems 
\cite{Bot84}. Hence, only for half-integer spins the $(1+1)$-dimensional 
D-theory with an infinite extent $\beta$ in the second direction has an 
infinite correlation length. The low-energy effective theory for this system is
the 2-d $O(3)$ model at vacuum angle $\theta=\pi$, i.e.,
\begin{eqnarray}
S[\vec s]&=&\int_0^\beta \! dx_1 \int \! dx_2 \ \{\frac{1}{2 g^2}
[\partial_1 \vec s \cdot \partial_1 \vec s + 
\frac{1}{c^2} \partial_2 \vec s \cdot \partial_2 \vec s] \nonumber \\
&+&\frac{i \theta}{4 \pi} \vec s \cdot 
[\partial_1 \vec s \times \partial_2 \vec s]\},
\end{eqnarray}
as conjectured by Haldane \cite{Hal83}. This model is in the universality class
of a 2-d conformal field theory --- the $k=1$ Wess-Zumino-Novikov-Witten model 
--- as was first argued by Affleck \cite{Aff88}.  Indeed, it has been shown 
numerically that the mass-gap of the 2-d $O(3)$ model (which is present at 
$\theta \neq \pi$) disappears at $\theta=\pi$ \cite{Bie95}. The simulation of 
the 2-d $O(3)$ model at $\theta = \pi$ is extremely difficult due to the 
complex action. Still, it is feasible using the Wolff cluster algorithm 
combined with an appropriate improved estimator for the topological charge 
distribution. It is remarkable that no complex action arises in the D-theory 
formulation of this problem, and the simulation is hence much simpler. 

So far, we have seen that D-theory naturally contains the continuum physics
of $O(3)$ models. This alone would be interesting. However, as we will see, 
D-theory is far more general. It can be extended to other scalar field theories
and to gauge theories with various symmetries and in various space-time 
dimensions.

\section{D-Theory Representation of Complex Vectors and Matrices}

As we have seen, in the low-temperature limit the 2-d spin 1/2 quantum 
Heisenberg model provides a D-theory regularization for the 2-d 
$O(3)$-symmetric continuum field theory. In that case, a vector of Pauli 
matrices replaces the 3-component unit-vector of a classical field 
configuration. Here, this structure is generalized to other fields. 

\subsection{Complex Vectors}

In $CP(N-1)$ models, classical $N$-component complex vectors $z$ arise. We will
now discuss their representation in D-theory. The symmetry group of a $CP(N-1)$
model is $U(N)$, which has $N^2$ generators. In D-theory the complex components
$z^i$ are represented by $2N$ Hermitean operators --- $N$ for the real and $N$ 
for the imaginary parts. Hence, the total number of generators is 
$N^2 + 2 N = (N+1)^2 - 1$ --- the number of generators of $SU(N+1)$. A rishon 
representation of the $SU(N+1)$ algebra is given by
\begin{equation}
Z^i = c^{0 \dagger} c^i, \ 
\vec G = \sum_{ij} c^{i \dagger} \vec \lambda_{ij} c^j, \ 
G = \sum_i c^{i \dagger} c^i.
\end{equation}
Here $\vec \lambda$ is the vector of generators for $SU(N)$, which obeys
$[\lambda^a,\lambda^b] = 2 i f_{abc} \lambda^c$, as well as 
$\mbox{Tr}(\lambda^a \lambda^b) = 2 \delta_{ab}$. The quantum operator $Z^i$ 
replaces the classical variable $z^i$, $\vec G$ is a vector of $SU(N)$ 
generators obeying $[G^a,G^b] = 2 i f_{abc} G^c$, and $G$ is a $U(1)$ 
generator.

\subsection{Complex Matrices}

$U(N)$ gauge theories are formulated in terms of classical complex matrix link
variables. The corresponding $SU(N)_L \otimes SU(N)_R \otimes U(1)$ gauge
transformations at the left- and right-hand side of a link are generated by 
$2(N^2-1)+1$ Hermitean operators. In D-theory a classical complex valued matrix
$u$ is replaced by a matrix $U$, whose elements are non-commuting operators. 
The elements of the matrix $U$ are described by $2N^2$ Hermitean generators ---
$N^2$ representing the real part and $N^2$ representing the imaginary part of 
the classical matrix $u$. Altogether, we thus have $2(N^2-1)+1+2N^2 = 4N^2-1$ 
generators --- the number of generators of $SU(2N)$. A rishon representation of
the $SU(2N)$ algebra is given by
\begin{eqnarray}
&&U^{ij} = c_+^{i \dagger} c_-^j, \nonumber \\ 
&&\vec L = \sum_{ij} c_+^{i \dagger} \vec \lambda_{ij} c_+^j, \
\vec R = \sum_{ij} c_-^{i \dagger} \vec \lambda_{ij} c_-^j, \nonumber \\ 
&&T = \sum_i (c_+^{i \dagger} c_+^i - c_-^{i \dagger} c_-^i).
\end{eqnarray}
There are two sets of rishons, $c_+^i$ and $c_-^i$, associated with the left 
and right $SU(N)$ symmetries generated by $\vec L$ and $\vec R$, and $T$ is a 
$U(1)$ generator. 

To summarize, in D-theory classical complex vectors $z$ are replaced by vectors
of operators $Z$, which are embedded in an $SU(N+1)$ algebra, and classical 
complex matrices $u$ are replaced by matrices $U$ with operator valued 
elements, which are embedded in the algebra of $SU(2N)$. Similarly, real valued
classical vectors and matrices can be replaced by corresponding operator valued
objects embedded in the algebras of $SO(N+1)$ and $SO(2N)$, respectively. In 
the next section we will use these basic structures to construct some D-theory 
models, both with global and local symmetries.

\section{D-theory Formulation of Specific Models}

In this section some quantum field theories are formulated in the framework of 
D-theory. Models with global and local symmetries are constructed explicitly. 
The Hamilton operator $H$ of a model is defined on a $d$-dimensional lattice. 
It replaces the Euclidean action in the standard formulation of lattice field 
theory. In D-theory, the partition function takes the quantum form 
$Z = \mbox{Tr} \exp(- \beta H)$. The dynamics of the models --- in particular, 
their dimensional reduction --- will be discussed in section 5.

\subsection{$U(N)$ and $SU(N)$ Quantum Link Models}

Wilson's formulation of lattice gauge theory uses classical complex $SU(N)$ 
parallel transporter link matrices $u_{x,\mu}$, with an action
\begin{equation}
S[u] = - \sum_{x,\mu \neq \nu} \mbox{Tr} [u_{x,\mu} u_{x+\hat\mu,\nu} 
u^\dagger_{x+\hat\nu,\mu} u^\dagger_{x,\nu}].
\end{equation}
The action is invariant under $SU(N)$ gauge transformations
\begin{equation}
u'_{x,\mu} = \exp(i \vec \alpha_x \cdot \vec \lambda) u_{x,\mu}
\exp(- i \vec \alpha_{x+\hat\mu} \cdot \vec \lambda).
\end{equation}
In D-theory the action is replaced by the Hamilton operator
\begin{equation}
\label{Hamiltonian}
H = J \sum_{x,\mu \neq \nu} \mbox{Tr} [U_{x,\mu} U_{x+\hat\mu,\nu} 
U^\dagger_{x+\hat\nu,\mu} U^\dagger_{x,\nu}].
\end{equation}
Here the elements of the $N \times N$ quantum link operators $U_{x,\mu}$ 
consist of generators of $SU(2N)$. Gauge invariance now means that $H$ commutes
with the local generators $\vec G_x$ of gauge transformations at the site $x$, 
which obey
\begin{equation}
[G^a_x,G^b_y] = 2 i \delta_{xy} f_{abc} G^c_x.
\end{equation}
Gauge covariance of a quantum link variable requires
\begin{eqnarray}
U'_{x,\mu}&=&\prod_y \exp(- i \vec \alpha_y \cdot \vec G_y) U_{x,\mu} 
\prod_z \exp(i \vec \alpha_z \cdot \vec G_z) \nonumber \\
&=&\exp(i \vec \alpha_x \cdot \vec \lambda) U_{x,\mu}
\exp(- i \vec \alpha_{x+\hat\mu} \cdot \vec \lambda),
\end{eqnarray}
where $\prod_x \exp(i \vec \alpha_x \cdot \vec G_x)$ is the unitary operator 
that represents a general gauge transformation in Hilbert space. The above 
equation implies the following commutation relation
\begin{equation}
\label{GUcommutator}
[\vec G_x,U_{y,\mu}] = \delta_{x,y+\hat\mu} U_{y,\mu} \vec \lambda -
\delta_{x,y} \vec \lambda U_{y,\mu}.
\end{equation}
It is straightforward to show that this is satisfied when we write
\begin{equation}
\vec G_x = \sum_\mu (\vec R_{x-\hat\mu,\mu} + \vec L_{x,\mu}),
\end{equation}
where $\vec R_{x,\mu}$ and $\vec L_{x,\mu}$ are generators of right and left 
gauge transformations of the link variable $U_{x,\mu}$. The commutation 
relations of eq.(\ref{GUcommutator}) imply
\begin{eqnarray}
&&[\vec R_{x,\mu},U_{y,\nu}] = 
\delta_{x,y} \delta_{\mu\nu} U_{x,\mu} \vec\lambda, \nonumber \\ 
&&[\vec L_{x,\mu},U_{y,\nu}] = - \delta_{x,y} \delta_{\mu\nu} \vec\lambda 
U_{x,\mu}.
\end{eqnarray}
The rishon representation is given by
\begin{eqnarray}
U^{ij}_{x,\mu}&=&c_{x,+\mu}^{i \dagger} c_{x+\hat\mu,-\mu}^j, \nonumber \\ 
\vec L_{x,\mu}&=&\sum_{ij} c_{x,+\mu}^{i \dagger} \vec \lambda_{ij} 
c_{x,+\mu}^j, \nonumber \\
\vec R_{x,\mu}&=&\sum_{ij} c_{x+\hat\mu,-\mu}^{i \dagger} \vec \lambda_{ij} 
c_{x+\hat\mu,-\mu}^j.
\end{eqnarray}
The rishon operators obey canonical anti-commutation relations
\begin{eqnarray}
\{c^i_{x,\pm\mu},c^{j \dagger}_{y,\pm\nu}\}&=&\delta_{xy} 
\delta_{\pm\mu,\pm\nu} \delta_{ij}, \nonumber \\ 
\{c^i_{x,\pm\mu},c^j_{y,\pm\nu}\}&=&0, \
\{c^{i \dagger}_{x,\pm\mu},c^{j \dagger}_{y,\pm\nu}\} = 0. 
\end{eqnarray}
The whole algebra commutes with the rishon 
number operator
\begin{equation}
{\cal N}_{x,\mu} = \sum_i (c^{i \dagger}_{x,+\mu} c^i_{x,+\mu} +
c^{i \dagger}_{x+\hat\mu,-\mu} c^i_{x+\hat\mu,-\mu})
\end{equation}
on each individual link. Together with the generator
\begin{equation} 
T_{x,\mu} = \sum_i (c_{x,+\mu}^{i \dagger} c_{x,+\mu}^i - 
c_{x+\hat\mu,-\mu}^{i \dagger} c_{x+\hat\mu,-\mu}^i).
\end{equation}
the above operators form the link based algebra of $SU(2N)$. One finds
\begin{equation}
[T_{x,\mu},U_{y,\nu}] = 2 \delta_{x,y} \delta_{\mu\nu} U_{x,\mu},
\end{equation}
which implies that
\begin{equation}
G_x = \frac{1}{2} \sum_\mu (T_{x-\hat\mu,\mu} - T_{x,\mu})
\end{equation}
generates an additional $U(1)$ gauge transformation, i.e.
\begin{eqnarray}
U'_{x,\mu}&=&\prod_y \exp(- i \alpha_y G_y) U_{x,\mu} \prod_z 
\exp(i \alpha_z G_z) \nonumber \\
&=&\exp(i \alpha_x) U_{x,\mu} \exp(- i \alpha_{x+\mu}).
\end{eqnarray}
Indeed the Hamilton operator of eq.(\ref{Hamiltonian}) is also invariant under 
the extra $U(1)$ gauge transformations and thus describes a $U(N)$ lattice 
gauge theory. 

\begin{figure}[t]
\epsfig{figure=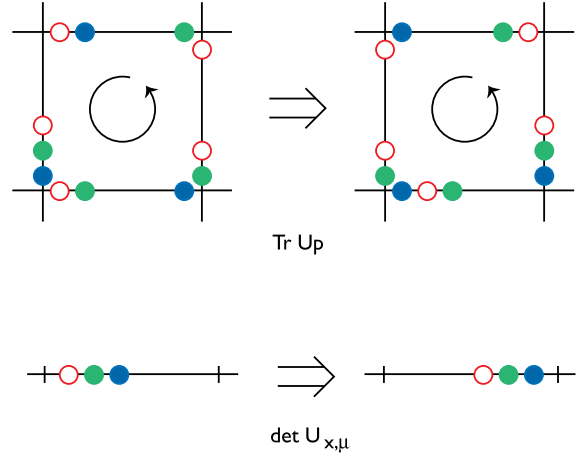,width=7.5cm}
\vspace*{-1cm}
\caption{QCD rishon dynamics: The trace part of the Hamiltonian induces hopping
of rishons of various colors around a plaquette. The determinant part shifts a 
color-neutral combination of $N$ rishons from one end of a link to the other.}
\vspace{-0.5cm}
\label{rishons}
\end{figure}

To reduce the symmetry of the quantum link model from $U(N)$ to $SU(N)$ one 
breaks the additional $U(1)$ gauge symmetry by adding the real part of the 
determinant of each link matrix to the Hamilton operator
\begin{eqnarray}
\label{Hamdet}
H&=&J \sum_{x,\mu \neq \nu} \mbox{Tr} [U_{x,\mu} U_{x+\hat\mu,\nu} 
U^\dagger_{x+\hat\nu,\mu} U^\dagger_{x,\nu}] \nonumber \\
&+&J' \sum_{x,\mu} \ [\mbox{det} U_{x,\mu} + \mbox{det} U^\dagger_{x,\mu}].
\end{eqnarray}
The $U(N)$ symmetry can be reduced to $SU(N)$ via the determinant, only when 
one works with ${\cal N}_{x,\mu} = N$ rishons on each link. This corresponds to
choosing the $(2N)!/(N!)^2$-dimensional representation of 
$SU(2N)$. The rishon dynamics is illustrated in figure \ref{rishons}.

\subsection{Quantum $CP(N-1)$ Models}

$CP(N-1)$ models are interesting, because they have a global $SU(N)$ symmetry 
as well as a $U(1)$ gauge invariance. However, the gauge fields are just 
auxiliary fields in this case. In the standard formulation of lattice field 
theory $CP(N-1)$ models are formulated in terms of classical complex 
unit-vectors $z_x$ and complex link variables $u_{x,\mu}$. The action can be 
written as
\begin{equation}
S[z,u] = - \sum_{x,\mu} (z^\dagger_x u_{x,\mu} z_{x+\hat\mu} +
z^\dagger_{x+\hat\mu} u^\dagger_{x,\mu} z_x).
\end{equation}
Note that there is no plaquette term for the gauge field. Consequently, the
gauge field can be integrated out --- it only acts as an auxiliary field. In
D-theory the corresponding Hamilton operator takes the form
\begin{equation}
H = J \sum_{x,\mu} (Z^\dagger_x U_{x,\mu} Z_{x+\hat\mu} +
Z^\dagger_{x+\hat\mu} U^\dagger_{x,\mu} Z_x).
\end{equation}
Here the $N$ components $Z^i_x$ of the quantum spin $Z_x$ consist of $2N$ 
Hermitean generators of $SU(N+1)$, and $U_{x,\mu} = S^1_{x,\mu} + 
i S^2_{x,\mu} = S^+_{x,\mu}$ is the raising operator of an $SU(2)$ algebra on 
each link. In rishon representation we have
\begin{equation}
Z^i_x = c^{0 \dagger}_x c^i_x, \ 
U_{x,\mu} = c^\dagger_{x,+\mu} c_{x+\hat\mu,-\mu}.
\end{equation}
Global $SU(N)$ transformations are generated by
\begin{equation}
\vec G = \sum_x \sum_{ij} c^{i \dagger}_x \vec \lambda_{ij} c^j_x,
\end{equation}
and the generator of $U(1)$ gauge transformations takes the form
\begin{equation}
G_x = \frac{1}{2} \sum_\mu (T_{x-\hat\mu,\mu} - T_{x,\mu}) + 
\sum_i c^{i \dagger}_x c^i_x.
\end{equation}
In rishon representation we have
\begin{equation}
T_{x,\mu} = c^\dagger_{x,+\mu} c_{x,+\mu} - 
c^\dagger_{x+\hat\mu,-\mu} c_{x+\hat\mu,-\mu}.
\end{equation}
The invariance properties of the model follow from $[H,\vec G] = [H,G_x] = 0$.
In this case, the rishon numbers on each site and on each link are separately
conserved.

It should be clear by now that D-theory is a very general structure, which
naturally provides us with lattice field theories formulated in terms of 
discrete variables --- quantum spins and quantum links. The cases worked out
here in some detail are examples that demonstrate the generality of D-theory. 
There are certainly more models one could investigate. Next, we want to argue
that D-theory does not define a new set of field theories. It just provides a 
new non-perturbative regularization and quantization of the corresponding 
classical models. To understand this, we must address the issue of dimensional
reduction.

\section{Classical Gauge Fields from Dimensional Reduction of Quantum Links}

As we have seen in detail in section 2, the quantum Heisenberg model provides
a D-theory regularization for the 2-d $O(3)$ model. The connection between the 
two models is established using chiral perturbation theory. The Goldstone 
boson fields $\vec s$ describing the low-energy dynamics of the Heisenberg 
model emerge as collective excitations of the quantum spins. By dimensional 
reduction they become the effective 2-d fields of a lattice $O(3)$ model. The 
continuum limit of this lattice model is reached when the extent $\beta$ of the
extra Euclidean dimension becomes large. The success of D-theory relies 
entirely on the fact that the $(d+1)$-dimensional theory is mass\-less, i.e., 
that in the ground state of the quantum Heisenberg model the $SO(3)$ symmetry 
is spontaneously broken to $SO(2)$. Only then chiral perturbation theory 
applies, and dimensional reduction occurs. The same is true for other D-theory 
quantum spin models. 

Let us now consider $SU(N)$ non-Abelian gauge theories in $d=4$. The 
Hamilton operator of the corresponding quantum link model, which is defined on 
a 4-d lattice, describes the evolution of the system in a fifth Euclidean
direction. The partition function $Z = \mbox{Tr} \exp(- \beta H)$ can then be
represented as a $(4+1)$-d path integral. Note that we have not included a
projector on gauge invariant states, i.e., gauge variant states also propagate
in the fifth direction. This means that we do not impose a Gauss law in the
unphysical direction. Not imposing Gauss' law implies $A_5 = 0$ for the fifth 
component of the gauge potential. This is important, because it leaves us with 
the correct field content after dimensional reduction. Of course, the physical 
Gauss law is properly imposed, because the model does contain non-trivial 
Polyakov loops in the Euclidean time direction. 

Dimensional reduction in quantum link models works differently than for quantum
spins. In the spin case the spontaneous breakdown of a global symmetry provides
the mass\-less Goldstone modes that are necessary for dimensional reduction. On
the other hand, when a gauge symmetry breaks spontaneously, the Higgs mechanism
gives mass to the gauge bosons, and dimensional reduction would not occur.
Fortunately, non-Abelian gauge theories in five dimensions are generically in
a mass\-less Coulomb phase \cite{Cre79}. This has recently been verified in 
detail for 5-d $SU(2)$ and $SU(3)$ lattice gauge theories using Wilson's 
formulation \cite{Bea98}. Here, we assume that the same is true for quantum 
link models --- an assumption that can only be checked in numerical simulations.
The leading terms in the low-energy effective action of 5-d Coulombic gluons 
take the form
\begin{eqnarray}
S[A]&=&\int_0^\beta dx_5 \int d^4x \ \frac{1}{2 e^2}[\mbox{Tr} \ 
F_{\mu\nu} F_{\mu\nu} \nonumber \\
&+&\frac{1}{c^2} \mbox{Tr} \ \partial_5 A_\mu \partial_5 A_\mu].
\end{eqnarray}
The quantum link model leads to a 5-d gauge theory characterized by the 
``velocity of light'' $c$. Note that here $\mu$ runs over 4-d indices only. The
dimensionful 5-d gauge coupling $1/e^2$ is the analog of $\rho_s$ in the spin
models. At finite $\beta$ the above theory has only a 4-d gauge invariance, 
because we have fixed $A_5 = 0$, i.e., we have not imposed the Gauss law. At
$\beta = \infty$ we are in the 5-d Coulomb phase with mass\-less gluons, and 
thus with an infinite correlation length $\xi$. When $\beta$ is made finite, 
the extent of the extra dimension is negligible compared to $\xi$. Hence, the 
theory appears to be dimensionally reduced to four dimensions. Of course, in 
four dimensions the confinement hypothesis suggests that gluons are no longer 
mass\-less. Indeed, as it was argued in ref.\cite{Cha97}, a finite correlation 
length
\begin{equation}
\xi \propto \exp(\frac{24 \pi^2 \beta}{11 N e^2})
\end{equation}
is expected to be generated non-perturbatively. Here $24 \pi^2/11 N$ is the
1-loop $\beta$-function coefficient of $SU(N)$ gauge theory. In contrast to the
spin models, the exact value for the mass-gap (and hence the prefactor of the
exponential) is not known in this case. For large $\beta$ the gauge coupling of
the dimensionally reduced 4-d theory is given by
\begin{equation}
1/g^2 = \beta/e^2.
\end{equation}
Thus the continuum limit $g \rightarrow 0$ of the 4-d theory is approached when
one sends the extent $\beta$ of the fifth direction to infinity. Hence,
dimensional reduction occurs when the extent of the fifth direction becomes 
large. This is due to asymptotic freedom, which implies that the correlation 
length grows exponentially with $\beta$. As in the spin models, it is useful to
think of the dimensionally reduced 4-d theory as a Wilsonian lattice theory 
with lattice spacing $\beta c$ (which has nothing to do with the lattice
spacing of the quantum link model). In fact, one can again imagine performing 
a block renormalization group transformation that averages the 5-d field over 
cubic blocks of size $\beta$ in the fifth direction and of size $\beta c$ in 
the four physical space-time directions. The block centers then form a 4-d 
space-time lattice of spacing $\beta c$ and the effective theory of the block 
averaged 5-d field is indeed an ordinary 4-d lattice theory.

The situation for 4-d $U(1)$ gauge theory is different. In contrast to 
non-Abelian gauge theories, after dimensional reduction from five to four 
dimensions, there is no reason for the photons to pick up a mass. They can 
exist in a 4-d Coulomb phase, because they are not confined. Hence, dimensional
reduction occurs already at a finite extent of the fifth dimension --- not only
in the $\beta \rightarrow \infty$ limit. Still, when $\beta$ becomes too small,
one enters the strong coupling confined phase, which has a finite correlation 
length. If the phase transition between the confined phase and the Coulomb 
phase is of second order, one obtains universal continuum behavior via 
dimensional reduction.

\section{Inclusion of Fermion Fields}

In order to formulate realistic quantum field theories in the framework of 
D-theory, it is important to include fermion fields. Even in Wilson's
formulation of lattice field theory, fermions already live in a finite Hilbert
space per site. Thus, their treatment can be taken over to D-theory with almost
no change. Here, we illustrate this for QCD. Other models with fermions can be
constructed along the same lines. In lattice gauge theory the fermion doubling 
problem arises. Wilson solved that problem by breaking chiral symmetry 
explicitly. The standard Wilson action for QCD, for example, is given by
\begin{eqnarray}
\label{Wilsonaction}
&&S[\bar\psi,\psi,u] = - \sum_{x,\mu \neq \nu} \mbox{Tr} 
[u_{x,\mu} u_{x+\hat\mu,\nu} u^\dagger_{x+\hat\nu,\mu} u^\dagger_{x,\nu}] 
\nonumber \\
&&+\frac{1}{2} \sum_{x,\mu} \ [\bar\psi_x \gamma_\mu u_{x,\mu} \psi_{x+\hat\mu}
- \bar\psi_{x+\hat\mu} \gamma_\mu u^\dagger_{x,\mu} \psi_x] \nonumber \\
&&+ M \sum_x \bar\psi_x \psi_x +\frac{r}{2} \sum_{x,\mu} \ 
[2 \bar\psi_x \psi_x \nonumber \\
&&-\bar\psi_x u_{x,\mu} \psi_{x+\hat\mu}
- \bar\psi_{x+\hat\mu} u^\dagger_{x,\mu} \psi_x].
\end{eqnarray}
Here $\bar\psi_x$ and $\psi_x$ are independent Grassmann valued spinors
associated with the lattice site $x$, $\gamma_\mu$ are Dirac matrices and
$M$ is the bare quark mass. The Wilson term (proportional to $r$) removes the
doubler fermions, but also breaks chiral symmetry. In practice, this leads to 
several problems, because $M$ must be tuned appropriately in order to reach the
chiral limit.

The construction principle of D-theory is to replace the classical action in
Wilson's formulation by a Hamilton operator that describes the evolution of the
system in an additional Euclidean direction. For fermions one must also replace
$\bar\psi_x$ by $\Psi^\dagger_x \gamma_5$. Hence, in D-theory, QCD is described
by the Hamilton operator
\begin{eqnarray}
\label{QCDaction}
H&=&J \sum_{x,\mu \neq \nu} \mbox{Tr} [U_{x,\mu} U_{x+\hat\mu,\nu} 
U^\dagger_{x+\hat\nu,\mu} U^\dagger_{x,\nu}] \nonumber \\
&+&J' \sum_{x,\mu} \ [\mbox{det} U_{x,\mu} + \mbox{det} U^\dagger_{x,\mu}] 
\nonumber \\
&+&\frac{1}{2} \! \sum_{x,\mu} [\Psi^\dagger_x \gamma_5 \gamma_\mu U_{x,\mu} 
\Psi_{x+\hat\mu}
- \Psi^\dagger_{x+\hat\mu} \gamma_5 \gamma_\mu U^\dagger_{x,\mu} \Psi_x] 
\nonumber \\
&+& M \sum_x \Psi^\dagger_x \gamma_5 \Psi_x + \frac{r}{2} \sum_{x,\mu} \ 
[2 \Psi^\dagger_x \gamma_5 \Psi_x \nonumber \\
&-&\Psi^\dagger_x \gamma_5 U_{x,\mu} \Psi_{x+\hat\mu}
- \Psi^\dagger_{x+\hat\mu} \gamma_5 U^\dagger_{x,\mu} \Psi_x].
\end{eqnarray}
Here $\Psi^\dagger_x$ and $\Psi_x$ are quark creation and annihilation 
operators with canonical anti-commutation relations
\begin{eqnarray}
&&\{\Psi^{i a \alpha}_x,\Psi^{j b \beta \dagger}_y\} = 
\delta_{xy} \delta_{ij} \delta_{ab} \delta_{\alpha \beta}, \nonumber \\
&&\{\Psi^{i a \alpha}_x,\Psi^{j b \beta}_y\} =
\{\Psi^{i a \alpha \dagger}_x,\Psi^{j b \beta \dagger}_y\} = 0,
\end{eqnarray}
where $(i,j)$, $(a,b)$ and $(\alpha,\beta)$ are color, flavor and Dirac 
indices, respectively. The generators of $SU(N)$ gauge transformation take the 
form
\begin{equation}
\vec G_x = \sum_\mu (\vec R_{x-\hat\mu,\mu} + \vec L_{x,\mu})
+ \Psi^\dagger_x \vec \lambda \Psi_x.
\end{equation}

The dimensional reduction of fermions is a subtle issue. If one would use
standard antiperiodic boundary conditions, familiar from thermodynamics, also 
in the extra dimension, the Matsubara modes, $p_5 = 2 \pi (n_5 + 
\frac{1}{2})/\beta$, would lead to a short $O(\beta)$ correlation length of the
dimensionally reduced fermion. On the other hand, the gluon dynamics happens at
distance scales which are exponentially large in $\beta$. As we have seen, 
$\beta$ acts as the lattice spacing of the dimensionally reduced effective 
Wilsonian theory. Fermions with antiperiodic boundary conditions in the fifth 
direction would hence remain at the cut-off and the dimensionally reduced 
theory would still be a Yang-Mills theory without quarks. Of course, there is 
no need to use antiperiodic boundary conditions for the fermions, because the
additional dimension is not Euclidean time. When one uses periodic boundary
conditions, a Matsubara mode, $p_5 = 0$, arises, and the quarks survive 
dimensional reduction. However, we then again face the fine-tuning problem of
Wilson fermions.

The fine-tuning  problem has been solved very elegantly in Shamir's variant 
\cite{Sha93} of Kaplan's fermion proposal \cite{Kap92}. Kaplan realized that
5-d fermions coupled to a 4-d domain wall develop a zero-mode bound to the 
wall. From the point of view of the dimensionally reduced theory, the zero-mode
represents a 4-d chiral fermion. In QCD, Shamir has used a simpler variant of 
Kaplan's method, which is formulated in a 5-d slab of finite size $\beta$ with 
open boundary conditions for the fermions at the two sides. This geometry fits 
naturally with the D-theory construction of quantum link models. The partition 
function with open boundary conditions for the quarks and with periodic 
boundary conditions for the gluons takes the form
\begin{equation}
Z = \mbox{Tr} \langle 0|\exp(- \beta H)|0\rangle.
\end{equation}
The trace extends over the gluonic Hilbert space only. Following Shamir, we
decompose the quark spinor into left and right-handed components
\begin{equation}
\Psi_x = \left( \begin{array}{c} \Psi_{Rx} \\ \Psi^\dagger_{Lx} \end{array} 
\right).
\end{equation}
Open boundary conditions for the fermions correspond to taking the expectation 
value of $\exp(- \beta H)$ in the Fock state $|0\rangle$, which is annihilated 
by the right-handed $\Psi_{R x}$ and by the left-handed $\Psi_{L x}$ 
\cite{Fur95}. Consequently, no left-handed quarks are present at the $x_5 = 0$
boundary, and no right-handed quarks exist at the $x_5 = \beta$ boundary of the
5-d slab.

In the presence of fermions, the low-energy effective theory of the gluons
(with $A_5 = 0$) is modified to
\begin{eqnarray}
\label{effaction}
&&S[\bar\psi,\psi,A] = \nonumber \\
&&\int_0^\beta \! dx_5 \int \! d^4x 
\{ \frac{1}{2 e^2} [\mbox{Tr} F_{\mu\nu} F_{\mu\nu} 
+ \frac{1}{c^2} \mbox{Tr} \partial_5 A_\mu \partial_5 A_\mu] \nonumber \\
&&+ \bar\psi[\gamma_\mu (A_\mu + \partial_\mu) + M + 
\frac{1}{c'} \gamma_5 \partial_5] \psi \}.
\end{eqnarray}
The ``velocity of light'' $c'$ of the quarks in the fifth direction is expected
to be different from the velocity $c$ of the gluons, because in D-theory there
is no symmetry between the four physical space-time directions and the extra 
fifth direction. This is no problem, because we are only interested in the 4-d 
physics after dimensional reduction.

Due to confinement, after dimensional reduction the gluonic correlation length 
is exponentially large, but not infinite. As explained in ref.\cite{Bro97}, the
same is true for the quarks, but for a different reason. Already free quarks 
pick up an exponentially small mass due to tunneling between the two 
boundaries of the 5-d slab. This mixes left- and right-handed states, and thus 
breaks chiral symmetry explicitly. The tunneling correlation length of the
quarks is given by $2 M \exp(M \beta)$. This allows an elegant solution of the 
notorious fine-tuning problem of the fermion mass $M$. In D-theory the gluon 
dynamics of quantum link QCD in the chiral limit takes place at a length scale
\begin{equation}
\xi \propto \exp(\frac{24 \pi^2 \beta}{(11 N - 2 N_f) e^2}),
\end{equation} 
which is determined by the 1-loop coefficient of the $\beta$-function of QCD 
with $N_f$ mass\-less quarks and by the 5-d gauge coupling $e$. Now one simply
chooses
\begin{equation}
M > \frac{24 \pi^2}{(11 N - 2 N_f) e^2},
\end{equation}
such that the chiral limit is reached automatically when one approaches the 
continuum limit by making $\beta$ large. This requires no fine-tuning.

\section{Conclusions}

We have seen that D-theory provides a rich structure, which allows us to 
formulate quantum field theories in terms of discrete variables --- quantum 
spins and quantum links. Dimensional reduction of discrete variables is a 
generic phenomenon. In $(d+1)$-dimensional quantum spin models with $d \geq 2$,
for example, it occurs because of spontaneous symmetry breaking, while in 
$(4+1)$-dimensional non-Abelian quantum link models it is due to the presence 
of a 5-d mass\-less Coulomb phase. The inclusion of fermions is very natural 
when one follows Shamir's variant of Kaplan's fermion proposal. In particular, 
the fine-tuning problem of Wilson fermions is solved very elegantly by going to
five dimensions. 

It is remarkable that D-theory treats bosons and fermions on an equal footing.
Both are formulated in a finite Hilbert space per site, both require the 
presence of an extra dimension, and both naturally have exponentially large
correlation lengths after dimensional reduction. The discrete nature of the
fundamental variables makes D-theory attractive, both from an analytic and from
a computational point of view. On the analytic side, the discrete variables
allow us to rewrite the theory in terms of fermionic rishon constituents of the
bosonic fields. This may turn out to be useful when one studies the large $N$ 
limit of various models. In particular, one can now carry over powerful 
techniques developed for condensed matter systems (like the quantum Heisenberg 
model) to particle physics. This includes the use of very efficient cluster 
algorithms, which has the potential of dramatically improving numerical 
simulations of lattice field theories. 

In D-theory the classical fields of ordinary quantum field theory arise via 
dimensional reduction of discrete variables. This requires specific dynamics 
--- namely a mass\-less theory in one more dimension. In general, the
verification of this basic dynamical ingredient of D-theory requires 
non-perturbative insight --- for example, via numerical simulations or via the 
large $N$ limit. Thus, the connection to ordinary field theory methods --- in
particular, to perturbation theory --- is rather indirect. This could be viewed
as a potential weakness, for example, because it seems hopeless to do 
perturbative QCD calculations in the framework of D-theory. However, the fact 
that perturbative calculations are virtually impossible may imply that 
non-perturbative calculations are now easier. After all, D-theory provides an 
additional non-perturbative microscopic structure underlying Wilson's lattice 
theory. Our hope is that this structure will help us to better understand the 
non-perturbative dynamics of quantum field theories.

\section*{Acknowledgments}

Quantum link models were developed together with S. Chandrasekharan, and 
extended to quantum link QCD and D-theory in collaboration with R. Brower. The 
condensed matter aspects of this work have been investigated together with 
B. B. Beard, R. J. Birgeneau, V. Chudnovsky and M. Greven. The other members of
the MIT D-theory collaboration, who have significantly contributed to the 
research presented here, are D. Chen, J. Cox, K. Holland, B. Scarlett, 
B. Schlittgen and A. Tsapalis. I thank all these people for a most enjoyable 
collaboration. The work described here is supported in part by funds provided  
by the U.S. Department of Energy (D.O.E.) under cooperative research agreement 
DE-FC02-94ER40818, as well as by the A. P. Sloan foundation.

\end{document}